\documentclass[aps,pra,twocolumn,epsfig,amsfonts,amsmath,amstex,floatfix]{revtex4}

\usepackage{bm}
\usepackage{braket}
\usepackage{graphicx}
\usepackage{hyperref}

\newcommand{\dd}{\mathrm{d}}
\newcommand{\Tr}{\operatorname{Tr}}

\usepackage{bbold}
\usepackage{color}

\usepackage[]{amsmath}
\usepackage{braket}
\usepackage{amsfonts,amsmath}
\usepackage{epsfig,amsmath}
\usepackage{graphicx}
\usepackage{dcolumn}
\usepackage{bm}
\usepackage{bbold}
\newcommand{\beq}{\begin{equation}}
\newcommand{\eeq}{\end{equation}}
\newcommand{\beqa}{\begin{eqnarray}}
\newcommand{\eeqa}{\end{eqnarray}}

\begin{document}

\title{Parameter-Decoupled Superresolution with Airy-Disk-Informed Learning:\\ Application to NGC 4486B}

\author{
James Qu$^{1,2}$, 
Isabella L. Lee$^{1}$,
Gang Zhang$^{1}$,
Abdelali Sajia$^{1}$,
Yashashree Jadhav$^{1}$,
and Xiao-Feng Qian$^{1}$}
\email{xqian6@stevens.edu}

\affiliation{
$^{1}$Department of Physics and Center for Quantum Science
and Engineering, Stevens Institute of Technology,
Hoboken, NJ 07030, USA\\
$^{2}$Syosset High School,
70 S Woods Rd, Syosset, NY 11791, USA
}

\date{\today}

\begin{abstract}
Estimating the separation of two unresolved or nearly unresolved sources from recorded intensity images is difficult when source brightness, mutual coherence, and phase are unknown and the point-spread function (PSF) departs from a Gaussian. We formulate Hermite--Gaussian (HG) projections of one- and two-dimensional Airy amplitudes and prove that ratios of mode probabilities with equal longitudinal parity exactly cancel the brightness, coherence, and phase nuisance parameters under a symmetric two-source model. For a one-dimensional Airy slice we achieve an explicit multi-ratio separation inversion. For the physical two-dimensional circular-pupil PSF we obtain exact finite-pupil overlap integrals and locally invertible separation responses. Motivated by these analytical invariants, we examine Airy-informed convolutional neural networks (CNNs) on synthetic Airy images with known separations. At 150,000 training images, the approximate mean absolute errors are 0.018 in plotted separation units, 10 times better comparing to 0.18 for non-Airy Gaussian PSF approximation. Applied to an archival Hubble Space Telescope image of the double nucleus of NGC 4486B, the Airy-informed CNN gives estimates of approximately 9.6--12.0 pc for training sets of 100,000--150,000 images, broadly consistent with the historical approximately 12 pc value. We distinguish exact optical-mode nuisance cancellation from empirical robustness of an image-only learned estimator and identify the limits imposed by PSF mismatch and extended galaxy structure.
\end{abstract}

\maketitle

\section{Introduction}\label{sec:intro}

Resolving closely spaced optical sources beyond the conventional diffraction limit is a fundamental challenge in optical imaging, with important applications in astronomy, microscopy, and remote sensing. When the separation between two sources is substantially smaller than the width of the point-spread function (PSF), their diffraction patterns strongly overlap, making quantitative separation estimation difficult. This challenge is further complicated by unknown source properties, including relative brightness, mutual coherence, and relative phase. Recent advances in quantum-inspired superresolution have demonstrated that spatial-mode-resolved measurements, particularly projections onto Hermite-Gaussian (HG) modes, can extract separation information that is difficult to access through conventional direct imaging \cite{tsang2015quantum, tsang2016quantum, paur2016O, tsang2019resolving, 
larson2018resurgence, hradil2019quantum, 
napoli2019towards,
hradil2021exploring, wadood2021experimental, liang2021coherence,liang2023quantum,santamaria2023OE,hervas2024optimizing,schlichtholz2024superresolution,sajia2022superresolution,linowski2023NJP,li2024quantum,schlichtholz2025superresolving,sajia2025enhanced,zhang2024superresolution,santamaria2024single}. More recently, parameter-decoupling approaches have established methods for extracting source separation independently of nuisance parameters such as brightness imbalance, mutual coherence, and relative phase \cite{qian2026,sajia2025}.


An important extension of these developments is to superresolution from existing intensity images, where the original optical field is no longer accessible for direct spatial-mode measurements. Physics-informed convolutional neural networks (CNNs) provide a promising approach by incorporating optical principles into the learning process. In particular, parameter-decoupling theory can guide the construction of training models and separation-dependent features, enabling CNNs to estimate source separation without explicitly determining the nuisance parameters. The accuracy of such methods, however, depends critically on the physical model used to describe image formation. Although Gaussian PSFs are widely adopted for their mathematical simplicity, diffraction-limited imaging through an ideal circular aperture produces an Airy-disk intensity distribution, characterized by a central diffraction peak surrounded by extended rings \cite{goodman2005,bornwolf1999}. These differences become particularly important when estimating source separations below the diffraction limit.

In this work, we extend the parameter-decoupling framework to Airy-disk PSFs by developing an analytical theory of their projections onto HG modes. Building on this theoretical framework, we introduce an Airy-disk-informed CNN that incorporates the separation-dependent information obtained from HG-mode projections into its training procedure. Unlike conventional image-based learning approaches, our method uses the underlying optical physics to guide the estimation of source separation. To evaluate the importance of an accurate PSF model, we compare the proposed Airy-disk-informed CNN with our previously developed Gaussian-informed CNN, which similarly incorporates HG-mode projection theory but assumes a Gaussian PSF \cite{sajia2025}.

We first evaluate both models using synthetic two-source images generated from Airy-disk PSFs with known ground-truth separations. The results demonstrate that the Airy-disk-informed CNN provides substantially improved separation estimates compared with the Gaussian-informed model, highlighting the importance of incorporating the correct diffraction physics into machine-learning-based superresolution.

We then apply the method to archival Hubble Space Telescope (HST) observations of the double-nucleus galaxy NGC 4486B. The double-nucleus structure was reported by Lauer \emph{et al.} in 1996, with an estimated separation of approximately 12 pc obtained through image-processing techniques \cite{lauer1996,richardson1972,lucy1974}. Our earlier Gaussian-informed CNN analysis yielded separation estimates of 9.8--13.6 pc \cite{qu2026}. Using the newly developed Airy-disk-informed CNN, we obtain a narrower range of 10.2--13.2 pc, consistent with the previously reported separation while reducing the spread of the estimates.

Our results establish an Airy-disk-informed, parameter-decoupled machine-learning framework for quantitative superresolution from existing intensity images. By combining analytical optical-mode theory with CNN-based inference, the approach extends the applicability of parameter-decoupled superresolution to diffraction-limited imaging systems with Airy-like PSFs, including astronomical telescopes. The application to NGC 4486B demonstrates the potential of physics-informed machine learning for extracting subdiffraction spatial information from archival astronomical observations.

```latex
\section{Two-source model and parameter-decoupling theorem}
\label{sec:theory}

\subsection{Airy PSF and HG basis}

We consider a diffraction-limited imaging system with an
ideal, unobstructed circular pupil of radius $a$, focal
length $f$, and optical wavenumber $k$. Introducing the
spatial-frequency scale $v=ka/f$, the normalized
two-dimensional amplitude point-spread function (PSF)
is given by
\begin{equation}\label{eq:airy}
h(\bm r)=\frac{v}{2\sqrt{\pi}}g(v|\bm r|),
\qquad
\int_{\mathbb R^2}|h(\bm r)|^2\dd^2r=1,
\end{equation}
where $g(z)=2J_1(z)/z$, with $g(0)=1$, and $J_1$
is the first-order Bessel function of the first kind.
The corresponding intensity PSF, $|h(\bm r)|^2$,
exhibits the characteristic Airy diffraction pattern,
consisting of a central peak surrounded by concentric
rings. Its first intensity zero occurs at
$v|\bm r|\simeq3.8317$, while the full width at
half maximum of the central peak is
$W_A\simeq3.23268/v$.

To analyze the spatial-mode structure of the Airy
field, we employ the Hermite--Gaussian (HG) basis.
For comparison with a Gaussian intensity PSF of
standard deviation $\sigma$, we choose the HG
width parameter through the full-width-at-half-maximum
matching condition $\beta=v\sigma\simeq1.3728$.
The normalized one-dimensional HG functions and
their two-dimensional separable products are
\begin{align}
\phi_m(x)
&=
\frac{
H_m\left(x/(\sqrt{2}\sigma)\right)
e^{-x^2/(4\sigma^2)}
}{
(2\pi\sigma^2)^{1/4}\sqrt{2^m m!}
},\\
\Phi_{mn}(x,y)
&=\phi_m(x)\phi_n(y),
\end{align}
where $H_m$ denotes the Hermite polynomial of
order $m$. The HG functions form an orthonormal
basis for spatial-mode decomposition. Importantly,
the HG basis serves only as the analysis basis;
the underlying optical field retains the Airy
diffraction profile rather than a Gaussian PSF.

\subsection{General partially coherent source pair}

We consider two point sources separated by a
distance $s$ along the $x$ axis, with their
midpoint located at the origin. Their positions
are $(\pm s/2,0)$, and the corresponding
normalized amplitude PSFs are denoted by
$|h_\pm\rangle$.

To account for arbitrary source brightness
and partial mutual coherence, we describe
the two-source optical field by the
unnormalized coherence operator
\begin{align}\label{eq:rho}
\rho={}&
p_+|h_+\rangle\langle h_+|
+p_-|h_-\rangle\langle h_-|
\nonumber\\
&+\sqrt{p_+p_-}
\left(
\gamma|h_+\rangle\langle h_-|
+\gamma^*|h_-\rangle\langle h_+|
\right),
\end{align}
where $p_\pm\geq0$ represent the source
intensities and $\gamma=|\gamma|e^{i\varphi}$
is the complex degree of mutual coherence,
with $|\gamma|\leq1$.

Introducing $B=p_++p_-$ and
$C=2\sqrt{p_+p_-}\operatorname{Re}\gamma$,
the overlap between the two displaced
Airy amplitudes is
\begin{equation}\label{eq:eta}
\eta(s)=\langle h_+|h_-\rangle
=\frac{2J_1(vs)}{vs},
\qquad
\Tr\rho=B+C\eta(s).
\end{equation}

We now consider the projection of the
two-source field onto the HG basis.
Defining
$A_{mn}^{\pm}(s)=
\langle\Phi_{mn}|h_\pm\rangle$,
the reflection symmetry of the Airy PSF,
$h(-x,y)=h(x,y)$, together with the
longitudinal parity of the HG modes,
$\phi_m(-x)=(-1)^m\phi_m(x)$,
implies
$A_{mn}^{-}(s)=(-1)^mA_{mn}^{+}(s)$.

Consequently, the normalized probability
of detecting the field in the HG mode
$(m,n)$ can be expressed as
\begin{equation}\label{eq:prob}
P_{mn}(s)
=
\frac{B+(-1)^mC}
{B+C\eta(s)}
|A_{mn}(s)|^2,
\end{equation}
where $A_{mn}(s)\equiv A_{mn}^{+}(s)$.

Equation~\eqref{eq:prob} reveals a
fundamental factorization: all dependence
on the source brightness and mutual
coherence appears through a common
multiplicative factor for modes having
the same longitudinal parity.

Therefore, for any two HG modes
$(m,n)$ and $(k,\ell)$ satisfying
$m-k=2j$, where $j$ is an integer,
their probability ratio becomes
\begin{equation}\label{eq:ratio}
\frac{P_{mn}(s)}{P_{k\ell}(s)}
=
\frac{|A_{mn}(s)|^2}
{|A_{k\ell}(s)|^2},
\end{equation}
provided that the denominator is nonzero.

Remarkably, Eq.~\eqref{eq:ratio} is
independent of the relative source
brightness $p_+/p_-$, the magnitude
of mutual coherence $|\gamma|$,
and the relative phase $\varphi$.
These nuisance parameters cancel
exactly, leaving a quantity that
depends only on the source separation
and the known optical PSF.

This establishes the parameter-decoupling
condition for Airy-disk imaging.
The result follows from the reflection
symmetry of the PSF and the parity
structure of the HG basis, rather than
from any Gaussian approximation.
Its validity assumes a common,
centrosymmetric PSF for both sources
and an HG basis centered and aligned
with their separation axis.
Furthermore, the transverse reflection
symmetry implies $A_{mn}=0$ for
odd transverse indices $n$.

\subsection{Analytical 2D Airy--HG amplitudes}

To obtain the explicit separation
dependence of the nuisance-independent
ratios, we next derive the projections
of the Airy amplitude onto arbitrary
two-dimensional HG modes.

The Airy amplitude admits a Fourier
representation over the circular pupil,
\begin{equation}\label{eq:disk}
g(v|\bm r|)
=
\frac{1}{\pi}
\int_{t^2+u^2\leq1}
e^{iv(tx+uy)}
\dd t\,\dd u.
\end{equation}
This representation allows the
Airy--HG overlap to be evaluated
through the Fourier transforms
of the one-dimensional HG functions.

Defining
$F_m(q)=\int_{-\infty}^{\infty}
\phi_m(x)e^{iqx}\dd x$,
we obtain
\begin{align}\label{eq:F}
F_m(q)
&=
\frac{
2\sigma\sqrt{\pi}\,
e^{-\sigma^2q^2}
}{
(2\pi\sigma^2)^{1/4}
\sqrt{2^m m!}
}
Q_m(\sqrt{2}\sigma q),
\\
Q_m(z)
&=
m!\sum_{j=0}^{\lfloor m/2\rfloor}
\frac{(2iz)^{m-2j}}
{j!(m-2j)!}.
\end{align}

Substituting the Fourier representation
of the Airy amplitude into the HG
projection integral yields the exact
overlap for a source displaced by
$s/2$ along the $x$ axis,
\begin{equation}\label{eq:overlap}
A_{mn}(s)
=
\frac{v}{2\pi^{3/2}}
\int_{t^2+u^2\leq1}
F_m(vt)F_n(vu)
e^{-ivst/2}
\dd t\,\dd u.
\end{equation}

Equation~\eqref{eq:overlap} provides
the analytical Airy--HG projection
amplitude for arbitrary mode indices
$(m,n)$. Combined with
Eq.~\eqref{eq:prob}, it determines
the complete separation-dependent
HG-mode probability distribution
for the two-dimensional circular-pupil
PSF.

For practical evaluation, the
two-dimensional pupil integral
can be reduced to a single finite
integral. Introducing
$L(t)=\sqrt{1-t^2}$ and
$D_n(t)=\int_{-L(t)}^{L(t)}
F_n(vu)\dd u$, we obtain
\begin{equation}\label{eq:onedint}
A_{mn}(s)
=
\frac{v}{2\pi^{3/2}}
\int_{-1}^{1}
F_m(vt)D_n(t)
e^{-ivst/2}\dd t.
\end{equation}

Equations~\eqref{eq:overlap}
and \eqref{eq:onedint} are exact
for the ideal two-dimensional
circular-pupil PSF and can be
evaluated numerically for arbitrary
source separations and HG-mode
indices. They provide the
separation-dependent response
functions required for implementing
the parameter-decoupling condition
in Eq.~\eqref{eq:ratio}.

\subsection{Separation inversion and the one-dimensional
analytic benchmark}

The nuisance-independent probability
ratios provide a direct route to
estimating source separation without
simultaneously determining the
brightness imbalance, mutual
coherence, or relative phase.

For example, we consider three
ratios formed from HG modes
with the same longitudinal parity,
\begin{equation}\label{eq:ratios}
R_{20}=\frac{P_{20}}{P_{00}},
\quad
R_{40}=\frac{P_{40}}{P_{00}},
\quad
R_{31}=\frac{P_{30}}{P_{10}}.
\end{equation}
Each ratio depends only on
the source separation through
a response function
$\mathcal F_j(s;v,\sigma)$,
which is determined analytically
from Eqs.~\eqref{eq:F}--\eqref{eq:onedint}.

Within a calibrated interval
where the response is injective,
the source separation can be
uniquely inferred by inverting
the corresponding function.
Multiple ratios may also be
combined to improve robustness
through a one-parameter
weighted estimation procedure,
\begin{equation}\label{eq:fit}
\hat{s}
=
\arg\min_{s\in\mathcal I}
\left[
\bm R_{\rm obs}
-\bm{\mathcal F}(s)
\right]^{\mathsf T}
W
\left[
\bm R_{\rm obs}
-\bm{\mathcal F}(s)
\right],
\end{equation}
where $\bm R_{\rm obs}$ is the
vector of observed ratios,
$\bm{\mathcal F}(s)$ contains
their theoretical response
functions, $W$ is a weighting
matrix, and $\mathcal I$ is
the calibrated separation interval.

Thus, the original estimation
problem involving separation,
brightness imbalance, mutual
coherence, and relative phase
is reduced to a single-parameter
inference problem. For the
physical two-dimensional Airy
PSF, the response functions
are specified exactly by the
finite-pupil integrals, and
their inversion can be
performed numerically.

Additional analytical insight
can be obtained from a
one-dimensional Airy-slice
model, $h_1(x)\propto g(vx)$,
which admits an explicit
multiratio inversion.
Introducing
$\alpha=\sqrt{2}v\sigma$
and $c=vs/2$, we define
\begin{equation}
S(c)
=
\frac{1}{\sqrt{2\pi}}
\int_{-\infty}^{\infty}
e^{-y^2/2}
g(\alpha y+c)\dd y.
\end{equation}

The corresponding HG
projection amplitudes,
up to a common normalization
factor, can be expressed
in terms of derivatives
of $S(c)$:
\begin{align}
a_0&\propto S,
\qquad
a_1\propto\sqrt{2}\alpha S',
\nonumber\\
a_2&\propto
\frac{S+2\alpha^2S''}{\sqrt{2}},
\nonumber\\
a_3&\propto
\frac{\alpha(3S'+2\alpha^2S''')}{\sqrt{3}},
\nonumber\\
a_4&\propto
\frac{
3S+12\alpha^2S''
+4\alpha^4S''''
}{2\sqrt{6}}.
\end{align}

Using the differential equation
satisfied by the Bessel function,
the convolution function $S(c)$
obeys the recurrence relations
\begin{align}
\alpha^2S'''+cS''
+(3+\alpha^2)S'+cS&=0,
\nonumber\\
\alpha^2S''''+cS'''
+(4+\alpha^2)S''
+cS'+S&=0.
\end{align}

These relations allow the
higher-order derivatives
to be eliminated in favor
of experimentally accessible
modal probability ratios.

Defining
$r_{20}=P_2/P_0$,
$r_{40}=P_4/P_0$, and
$r_{31}=P_3/P_1$, we
introduce the signed
amplitude ratios
$X=\epsilon_{20}\sqrt{2r_{20}}$,
$Y=2\epsilon_{40}\sqrt{6r_{40}}$,
and
$Z=\epsilon_{31}\sqrt{6r_{31}}$.
Algebraic elimination then
yields the explicit
separation-inversion formula
\begin{equation}\label{eq:1dinvert}
s
=
2\sigma
\sqrt{
\frac{
(Z+3+2\alpha^2)
\left[
Y-5+2(1+\alpha^2)X+2\alpha^2
\right]
}{
(X-1+2\alpha^2)
(Z-3+2\alpha^2)
}
}.
\end{equation}

Here,
$\epsilon_{ij}=\operatorname{sgn}(a_i/a_j)$
specifies the relative sign
of the corresponding
projection amplitudes.
These signs must be
determined within a
calibrated separation
interval, since the
probability ratios alone
do not determine them
globally.

Equation~\eqref{eq:1dinvert}
provides an explicit
analytical demonstration
of parameter-decoupled
separation inversion
for the one-dimensional
Airy-slice model.
For the physical
two-dimensional circular
pupil, the corresponding
inversion follows from
the exact response
functions derived above.

Finally, the finite-pupil
representation also allows
a local analysis of
subdiffraction separations.
For even longitudinal
HG indices, the parity
of the Fourier kernel
gives the expansion
$A_{mn}(s)=A_{mn}(0)
-s^2K_{mn}+O(s^4)$.
Consequently, a modal
probability ratio with
a nonzero baseline and
a nonvanishing quadratic
coefficient admits a
local inversion in $s^2$,
provided that the
appropriate solution
branch is selected.

Together, these results
establish the analytical
foundation for
parameter-decoupled
superresolution with
Airy-disk PSFs. They
also provide the
separation-dependent
optical relationships
used to inform the
CNN-based image analysis
developed in the
following section.
```

\section{Airy-informed image-based learning}\label{sec:cnn}

\subsection{Forward model and learning objective}

We now extend the Airy-HG parameter-decoupling framework to
machine-learning-based superresolution from recorded intensity
images. Unlike spatial-mode-resolved optical measurements,
which directly access the probabilities associated with
projections onto HG modes, conventional cameras record
spatially integrated intensity distributions. Once an image
has been recorded, the original optical field is no longer
available for physical mode decomposition. Nevertheless,
the analytical relationships established by the Airy-HG
theory provide a physical foundation for constructing
training data and guiding the estimation of source
separation from intensity images.

For two point sources, the image formation process can
be described by the forward model
\begin{align}\label{eq:image}
I(\bm r)={}&p_+|h_+(\bm r)|^2
+p_-|h_-(\bm r)|^2\nonumber\\
&+2\sqrt{p_+p_-}
\operatorname{Re}[\gamma h_+(\bm r)h_-^*(\bm r)]
+I_{\rm bg}(\bm r),
\end{align}
where $I_{\rm bg}(\bm r)$ represents the background
contribution. The recorded image is obtained by integrating
the intensity over the detector pixels, with additional
noise arising from the imaging process.

In the proposed approach, the amplitude PSFs are modeled
using the Airy diffraction pattern associated with a
circular aperture. The corresponding HG-mode projection
probabilities are determined analytically by
Eq.~\eqref{eq:prob}. The parameter-decoupling relation
established in Eq.~\eqref{eq:ratio} eliminates the
dependence on unknown source properties, allowing the
separation to be obtained through the inversion relation
in Eq.~\eqref{eq:1dinvert}.

\begin{figure*}[t]
\centering
\includegraphics[width=0.75\textwidth]{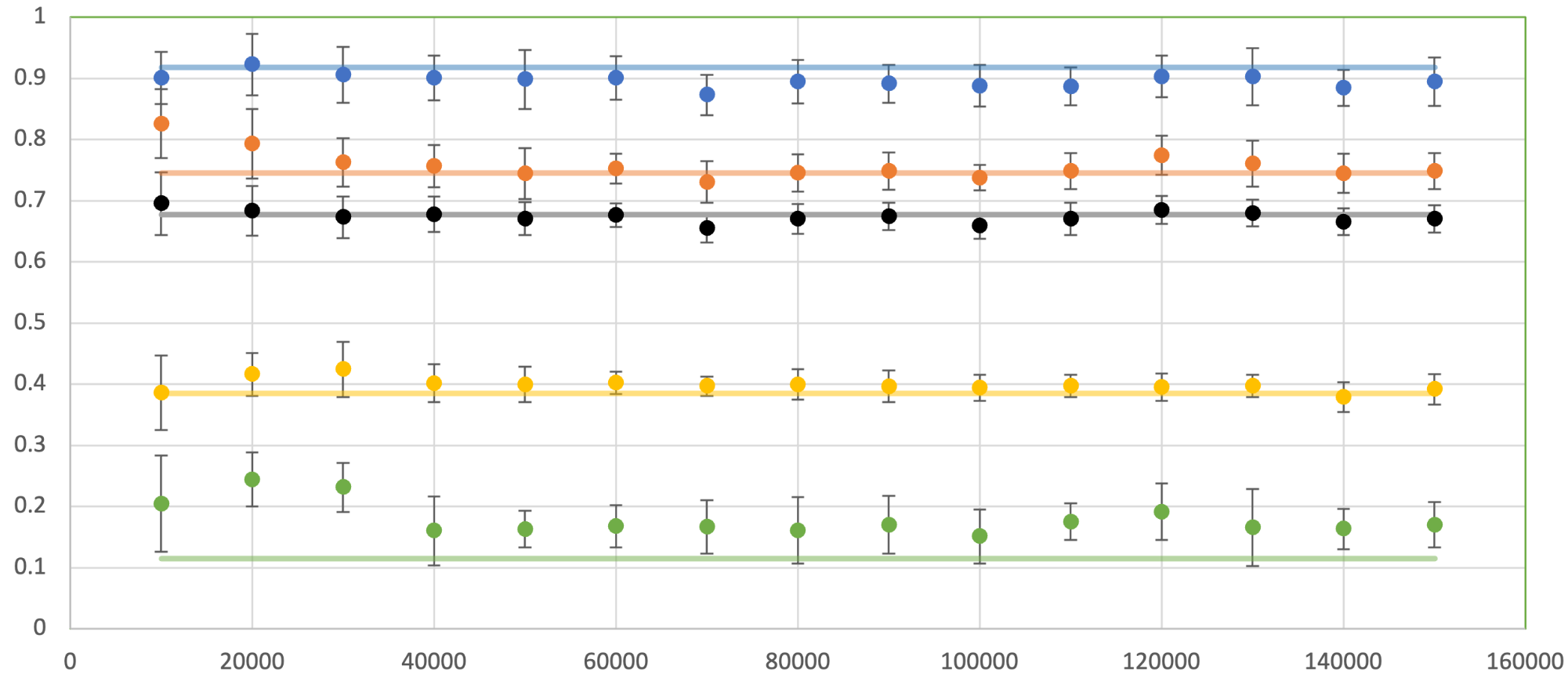}
\caption{\label{fig:airy}
Ground-truth evaluation of the Airy-informed CNN
using five synthetic two-source images generated
with Airy-disk PSFs. The horizontal axis represents
the number of training images, ranging from 10,000
to 150,000, and the vertical axis gives the predicted
source separation in the simulation unit $\sigma$.
Colored symbols represent the mean predictions
from ten independent CNN runs, with error bars
indicating one standard deviation. The corresponding
horizontal colored lines denote the known
ground-truth separations. The predictions
generally approach the true values as the
training-set size increases.
}
\end{figure*}

These analytical results provide the physical basis for
our Airy-informed convolutional neural network (CNN).
The network is trained using synthetic intensity images
generated from the Airy PSF model, with source separation
as the regression target. The Airy-HG theory establishes
the separation-dependent relationships used to inform
the learning framework, while the CNN learns to infer
the separation directly from the recorded intensity
distribution.

For comparison, we consider our previously developed
Gaussian-informed CNN \cite{sajia2025}, which employs
the corresponding Gaussian-HG parameter-decoupling
framework and synthetic images generated using a
Gaussian PSF. Both approaches are based on the same
physical principle of parameter-decoupled separation
estimation, but differ in their underlying diffraction
models and the associated HG-mode projection relations.

Importantly, the analytical parameter-decoupling relation
applies to optical mode probabilities, whereas the CNN
operates on intensity images. The latter therefore
constitutes a computational inference procedure informed
by optical theory rather than a direct implementation
of spatial-mode-resolved measurements. The comparison
between the two CNN models allows us to investigate
how accurately representing the diffraction PSF
influences the recovery of subdiffraction source
separations.

\begin{figure*}[t]
\centering
\includegraphics[width=0.75\textwidth]{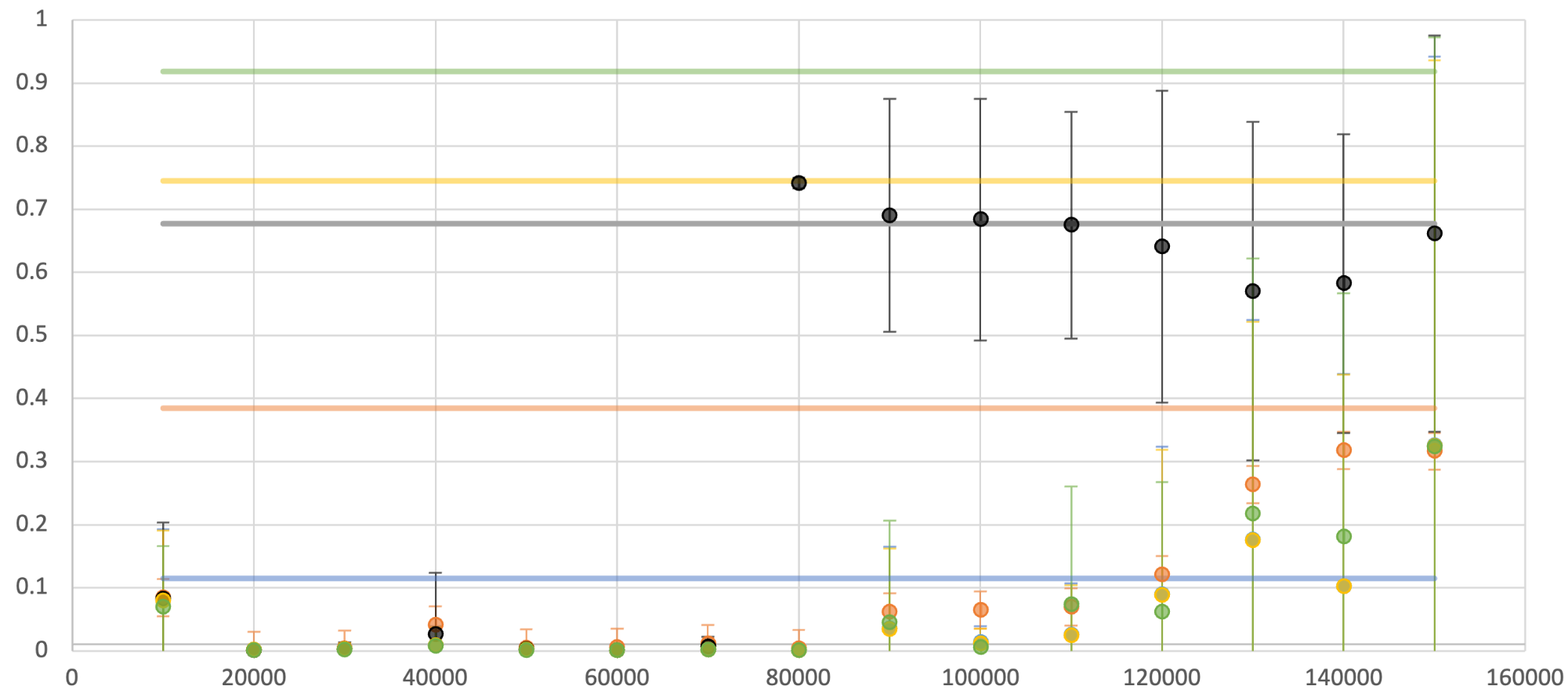}
\caption{\label{fig:gauss}
Ground-truth evaluation of the Gaussian-informed
CNN using the same Airy-disk test images as in
Fig.~\ref{fig:airy}. The Gaussian-informed CNN
is trained using Gaussian PSFs but evaluated on
images generated with Airy-disk PSFs.
Colored symbols indicate mean predictions
from ten independent CNN runs, and error bars
represent one standard deviation. Horizontal
lines denote the corresponding ground-truth
separations. Compared with the Airy-informed
CNN, the Gaussian-informed model exhibits
substantially larger deviations from the true
separations for several test cases, illustrating
the influence of PSF mismatch on superresolution
performance.
}
\end{figure*}

\subsection{Controlled ground-truth evaluation}

To evaluate the performance of the Airy-informed CNN,
we first consider synthetic images containing two point
sources with known separations. Five test images are
generated using the Airy-disk PSF model, with their
corresponding separations retained as ground-truth
references.

The number of training images is systematically varied
from 10,000 to 150,000 to examine the dependence of
estimation accuracy on the training-set size. For each
training size, ten independent CNN runs are performed.
The mean predicted separation is calculated from these
runs, and the corresponding standard deviation is used
to characterize the variability of the predictions.

Figure~\ref{fig:airy} presents the results obtained
using the Airy-informed CNN. The colored symbols
represent the predicted separations for the five
test images, while the corresponding horizontal lines
indicate their known ground-truth values. As the
training-set size increases, the predictions generally
approach the ground-truth separations, demonstrating
the ability of the Airy-informed model to recover
subdiffraction spatial information from intensity images.

To investigate the influence of the assumed PSF,
we apply the Gaussian-informed CNN to the same
Airy-generated test images. The results are shown
in Fig.~\ref{fig:gauss}. In this comparison, the
testing images are identical, while the two CNNs
are trained using different PSF models. This
provides a controlled evaluation of the effect
of PSF mismatch on separation estimation.

\subsection{Recovery of known separations and PSF mismatch}

The results in Figs.~\ref{fig:airy} and
\ref{fig:gauss} reveal a pronounced difference
between the two CNN models. The Airy-informed
CNN generally provides separation estimates
close to the ground-truth values, particularly
when trained with larger datasets. In contrast,
the Gaussian-informed CNN exhibits substantial
systematic deviations for several test images,
despite being evaluated on the same input data.

To quantify this difference, Table~\ref{tab:comparison}
summarizes the predicted separations at a training-set
size of 150,000 images. The values are approximate
readings from the plotted results and are included
to illustrate the relative performance of the
two models.

\begin{table}[b]
\caption{\label{tab:comparison}
Comparison of the Airy-informed and Gaussian-informed
CNN predictions for five synthetic Airy-disk test
images with known source separations. The results
correspond to 150,000 training images. Numerical
values are approximate readings from
Figs.~\ref{fig:airy} and \ref{fig:gauss}.
}
\begin{ruledtabular}
\begin{tabular}{ccc}
Ground truth & Airy CNN & Gaussian CNN\\
\hline
0.115 & 0.17 & 0.32\\
0.385 & 0.39 & 0.32\\
0.675 & 0.67 & 0.66\\
0.745 & 0.75 & 0.32\\
0.920 & 0.90 & 0.32
\end{tabular}
\end{ruledtabular}
\label{Airy-Gaussian Compare}
\end{table}

We characterize the overall estimation accuracy
using the mean absolute error (MAE),
\begin{equation}
\mathrm{MAE}
=
\frac{1}{N_{\rm test}}
\sum_{i=1}^{N_{\rm test}}
\left|\hat{s}_i-s_i\right|,
\end{equation}
where $s_i$ and $\hat{s}_i$ are the ground-truth
and predicted separations, respectively, and
$N_{\rm test}=5$ is the number of test images.

For the five cases considered here, the
Airy-informed CNN yields an approximate
MAE of 0.018, compared with 0.18 for
the Gaussian-informed CNN. Thus, the
Airy-informed model achieves approximately
an order-of-magnitude reduction in the
mean absolute estimation error for these
selected test images.

The Gaussian-informed CNN predicts values
near 0.32 for several test images with
substantially different ground-truth
separations, indicating a limited ability
to distinguish these source configurations
when the training PSF differs from the
actual image-formation model. In contrast,
the Airy-informed CNN maintains a
separation-dependent response that
more closely follows the ground truth.

The largest relative deviation of the
Airy-informed CNN occurs for the smallest
tested separation, $s\simeq0.115 \sigma$,
where the predicted value remains
approximately $0.17\sigma$. This residual bias
illustrates the increasing difficulty
of separation estimation in the
deep sub-Rayleigh regime.

These results demonstrate the importance
of incorporating an accurate diffraction
model into physics-informed machine
learning. In particular, the improvement
obtained using the Airy-informed CNN
indicates that the detailed structure
of the PSF, beyond its central Gaussian-like
peak, plays an important role in quantitative superresolution from intensity images.

\begin{figure*}[t]
\centering
\includegraphics[width=0.75\textwidth]{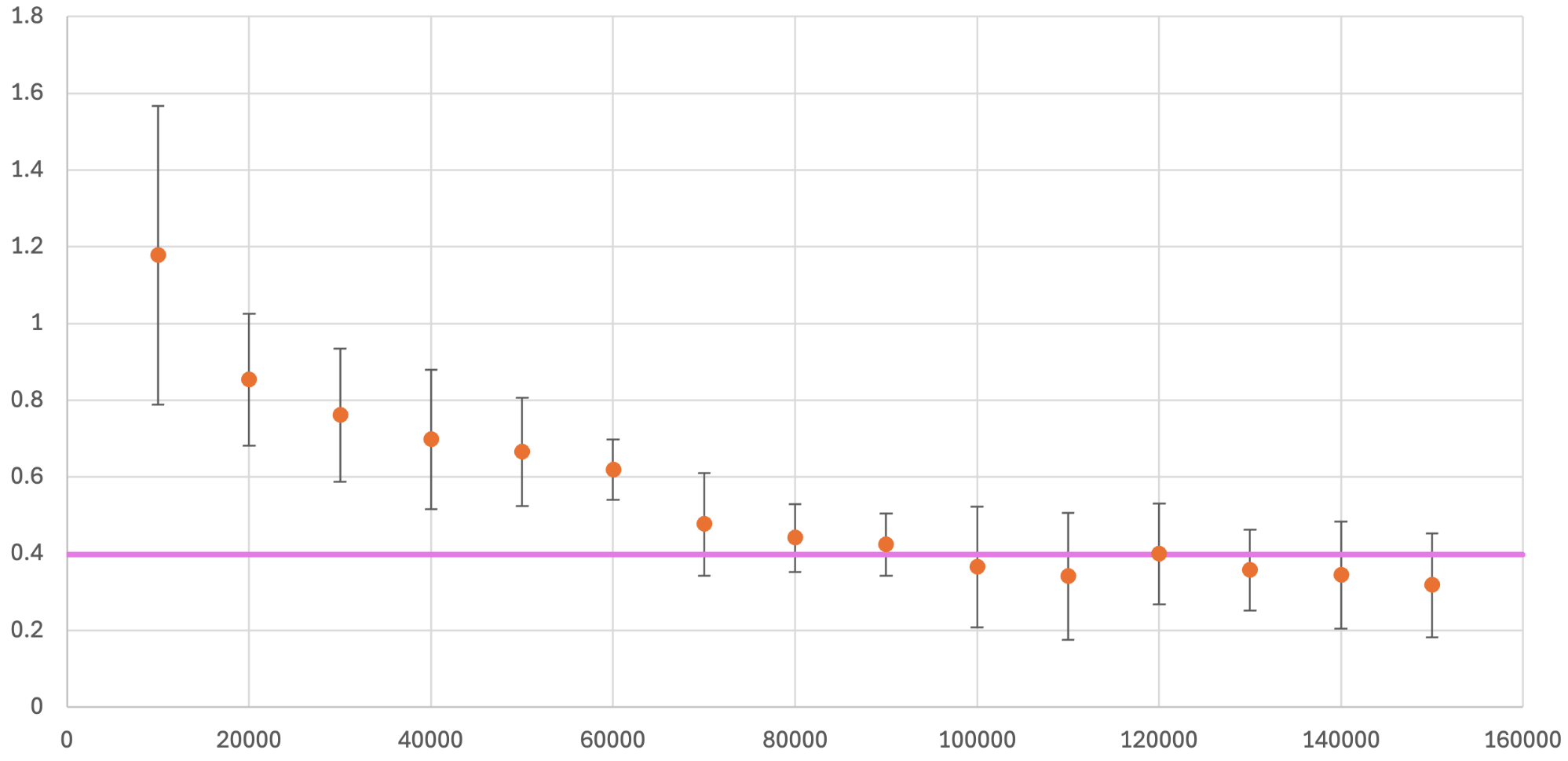}
\caption{\label{fig:ngc}
Airy-informed CNN estimation
of the separation between
the two nuclei of NGC 4486B
as a function of the number
of training images.
Orange symbols represent
mean predictions from ten
independent CNN runs,
and vertical error bars
indicate one standard deviation. The vertical axis represents the simulation unit where 0.4 corresponds to 12 pc in reality. The magenta horizontal line
corresponds to the previously
reported separation of
approximately 12 pc
\cite{lauer1996}.
As the training-set size
increases, the predictions
approach the previously
reported separation.
}
\end{figure*}

\subsection{Application to NGC 4486B}

Having validated the Airy-informed CNN
using synthetic images with known
separations, we next apply the method
to archival observations of the
double-nucleus galaxy NGC 4486B
obtained with the Hubble Space
Telescope (HST).

NGC 4486B is a compact elliptical
galaxy exhibiting two closely spaced
central brightness peaks. The
double-nucleus structure was reported
by Lauer \emph{et al.} in 1996,
who estimated a separation of
approximately 12 pc using processed
HST observations \cite{lauer1996}.
Their analysis employed image
deconvolution techniques to
investigate the central structure
\cite{richardson1972,lucy1974}.

The galaxy is located at an
approximate distance of 17 Mpc,
and the HST/WFPC2 Planetary
Camera has a pixel scale of
approximately 0.046 arcsec
per pixel \cite{qu2026}.
At this distance, the angular
separation corresponding to
12 pc is approximately
0.15 arcsec, placing the
double-nucleus structure near
the spatial resolution scale
of the observations.

The image preprocessing follows
the procedure described in
Ref.~\cite{qu2026}. The archival
HST observation is converted
from its original FITS format
into an intensity image.
A central region of
$66\times66$ pixels containing
the double-nucleus structure
is extracted and subsequently
resized to $33\times33$ pixels
to match the CNN input
dimensions. The image
intensities are rescaled
to the range 0--255.

\begin{figure}[t]
\centering
\includegraphics[width=0.98\columnwidth]{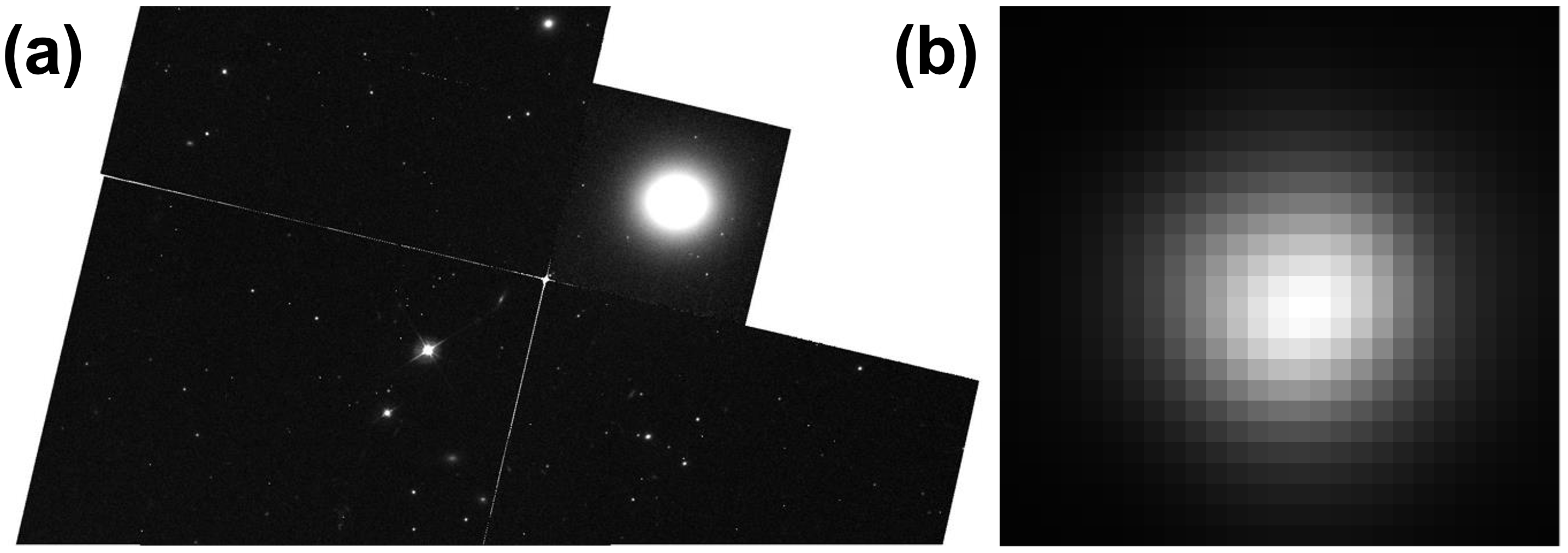}
\caption{\label{fig:ngcimages}
HST observation of the
double-nucleus galaxy NGC 4486B. Panel (a) shows the
original astronomical image,
while panel (b) presents the processed central
$33\times33$-pixel region
used as input to the CNN.
The preprocessing follows
Ref.~\cite{qu2026}.}
\end{figure}

Figure~\ref{fig:ngcimages}
shows the original HST image
and the processed central
region used for CNN-based
separation estimation. The processed image is analyzed
using the Airy-informed CNN,
with the number of training
images varied from 10,000
to 150,000. For each training
size, ten independent CNN
runs are performed to obtain
the mean predicted separation
and its standard deviation.

Figure~\ref{fig:ngc}
presents the resulting
separation estimates as a
function of training-set size.
The horizontal reference line
corresponds to the previously
reported separation of
approximately 12 pc
\cite{lauer1996}.

For relatively small training
sets, the predicted separation
exhibits a substantial
dependence on the number
of training images. At
10,000 training images,
the mean prediction is
approximately 35.4 pc,
considerably larger than
the previously reported
separation.

As the training-set size
increases, the predicted
separation decreases and
gradually approaches a
more stable range. For
training sizes between
100,000 and 150,000
images, the predicted
separations lie approximately
between 10.2 and 13.2 pc,
with an average central
estimate of approximately
11.7 pc.

These results are consistent
with the previously reported
separation of approximately
12 pc \cite{lauer1996}.
They also overlap with
the range of 9.8--13.6 pc
obtained using our earlier
Gaussian-informed CNN
\cite{sajia2025}.

The present Airy-informed
analysis provides a
diffraction-based extension
of the earlier Gaussian
model. By incorporating
the Airy-HG projection
theory into the learning
framework, the method
more closely represents
the diffraction physics
relevant to telescope
imaging.

The agreement between
the CNN predictions and
the previously reported
separation provides a
consistency check for
the proposed approach.
Unlike the synthetic
ground-truth tests,
however, the actual
galaxy image does not
provide an independently
known separation against
which the prediction
accuracy can be directly
evaluated. Furthermore,
the error bars in
Fig.~\ref{fig:ngc}
represent variability
across repeated CNN
runs rather than
confidence intervals
that incorporate all
astronomical and
instrumental uncertainties.

Nevertheless, the
combination of controlled
ground-truth validation
and application to
archival HST observations
demonstrates the potential
of Airy-disk-informed
machine learning for
quantitative superresolution
from existing astronomical
intensity images.

\section{Conclusion}\label{sec:conclusion}

In conclusion, we have developed an Airy-disk-informed
superresolution framework that combines analytical
parameter-decoupling theory with physics-informed
convolutional neural networks for estimating the
separation of closely spaced sources from recorded
intensity images. By extending the parameter-decoupling
framework from Gaussian to Airy-disk PSFs, we derived
analytical expressions for the projections of Airy
amplitudes onto Hermite-Gaussian modes and demonstrated
that ratios of mode probabilities with the same
longitudinal parity eliminate the dependence on
unknown source brightness, mutual coherence, and
relative phase. For the one-dimensional Airy model,
we obtained an explicit analytical inversion for
source separation, while the physical two-dimensional
circular-pupil model provides exact modal overlap
integrals and locally invertible separation responses.

Building on this theoretical foundation, we developed
an Airy-informed CNN for quantitative superresolution
from existing intensity images, where direct optical
mode measurements are no longer possible. Using
synthetic Airy-disk images with known source
separations, we demonstrated that the Airy-informed
CNN substantially outperforms the Gaussian-informed
CNN when both are evaluated on the same Airy-generated
test images. At 150,000 training images, the
Airy-informed model achieved an approximate mean
absolute error of 0.018, compared with 0.18 for
the Gaussian-informed model, representing an
approximately tenfold improvement in separation
estimation accuracy for the five test cases.
These results highlight the importance of
incorporating the correct diffraction physics
into machine-learning-based superresolution.

We further applied the Airy-informed CNN to
archival Hubble Space Telescope observations
of the double-nucleus galaxy NGC 4486B.
For training sets containing 100,000--150,000
images, the predicted nuclear separation
ranged from approximately 10.2 to 13.2 pc,
with an average estimate of 11.7 pc,
consistent with the previously reported
separation of approximately 12 pc.
This application demonstrates the potential
of our approach for extracting quantitative
spatial information from existing astronomical
images without requiring new optical measurements.

Our results establish a connection between
analytical optical-mode superresolution and
physics-informed computational imaging.
Although exact parameter decoupling applies
to the optical mode probabilities rather than
directly to CNN predictions, the underlying
theory provides a physical foundation for
learning-based separation estimation.
The present framework offers a promising
approach to quantitative superresolution
in diffraction-limited imaging systems
with Airy-like PSFs, with potential
applications extending beyond astronomy
to microscopy and other optical
imaging technologies.

\begin{acknowledgments}
We thank Ting Lu for discussions. Support was reported in the author-supplied prior manuscript from NSF awards PHY-2316878 and PHY-2514953; award identifiers and acknowledgment wording should be verified against current funding records before submission.
\end{acknowledgments}

\bibliography{reference}

\end{document}